\documentclass[journal = nalefd, manuscript=article]{achemso}
\usepackage{setspace}
\usepackage{graphicx}
\usepackage{amsmath}
\usepackage{lineno}
\usepackage[ngerman,english]{babel}
\usepackage{soul}

\title{Plasmon-Exciton Coupling Using DNA Templates}
\author{Eva-Maria Roller}
\affiliation{Faculty of Physics and Center for NanoScience (CeNS), Ludwig-Maximilians-Universit\"at (LMU), Munich 80539, Germany}
\author{Christos Argyropoulos}
\affiliation{Department of Electrical and Computer Engineering, University of Nebraska-Lincoln, Lincoln, Nebraska 68588, USA}
\author{Alexander H\"ogele}
\affiliation{Faculty of Physics and Center for NanoScience (CeNS), Ludwig-Maximilians-Universit\"at (LMU), Munich 80539, Germany}
\author{Tim Liedl}
\affiliation{Faculty of Physics and Center for NanoScience (CeNS), Ludwig-Maximilians-Universit\"at (LMU), Munich 80539, Germany}
\author{Mauricio Pilo-Pais}
\affiliation{Faculty of Physics and Center for NanoScience (CeNS), Ludwig-Maximilians-Universit\"at (LMU), Munich 80539, Germany}
\email{m.pilopais@lmu.de}

\begin{document}

\vspace{-35pt}
\maketitle

\begin{abstract}
Coherent energy exchange between plasmons and excitons is a phenomenon that arises in the strong coupling regime resulting in distinct hybrid states. The DNA-origami technique provides an ideal framework to custom-tune plasmon-exciton nanostructures. By employing this well controlled self-assembly process, we realized hybrid states by precisely positioning metallic nanoparticles in a defined spatial arrangement with fixed nanometer-sized interparticle spacing. Varying the nanoparticle diameter between $30\,nm$ and $60\,nm$ while keeping their separation distance constant allowed us to precisely adjust the plasmon resonance of the structure to accurately match the energy frequency of a J-aggregate exciton. With this system we obtained strong plasmon-exciton coupling and studied far-field scattering at the single-structure level. The individual structures displayed normal mode splitting up to $170\,meV$. The plasmon tunability and the strong field confinement attained with nanodimers on DNA-origami renders an ideal tool to bottom-up assembly plasmon-exciton systems operating at room temperature.
\end{abstract}


Nanoparticles (NPs) subjected to light  excitation exhibit collective oscillations of electrons (\textit{plasmons}), which in turn can greatly affect the behavior of quantum emitters positioned in nearby locations. The resulting plasmon-exciton coupling is of interest as it may facilitate studies of fundamental quantum phenomena such as coherent energy exchange, entanglement, and cavity quantum electrodynamics.~\cite{Tame2013} Potential applications of strongly coupled-exciton systems include artificial light harvesting,~\cite{Gonzalez-Ballestero2015} threshold-less lasing, or their use in quantum information processing.~\cite{Torma2015}  
The degree of interaction between plasmons and quantum emitters can be classified based on their coupling strength ($g$), displaying different signatures in the far-field scattering spectra, such as enhanced absorption dip, Fano resonance, or Rabi splitting.~\cite{Torma2015} Although these effects are usually associated with quantum-mechanical phenomena, they can be qualitatively described by classical electrodynamics.~\cite{Savasta2010, Faucheaux2014, Torma2015}
Plasmon frequencies can be tuned by varying the metallic NP size, geometry, interparticle separation, and their two- or three dimensional arrangement. Moreover, near-field enhancement can be obtained using small gaps among metallic NPs or using structures with sharp morphology.~\cite{Novotny2011} If an exciton is placed in regions with enough field confinement, it is possible to achieve the necessary coupling strength to reach the regime of strong coupling, which results in a normal mode splitting, in close analogy to a coupled harmonic oscillator.~\cite{Torma2015} Our focus is on the strong coupling regime where the energy exchange between the plasmon and the exciton results in distinct hybrid modes, the so-called \textit{plexciton} states. \\
Experimental realizations of plasmon-exciton coupling include work on metallic films,~\cite{Bellessa2004} lithographic constructs,~\cite{Bellessa2009, Schlather2013} and individual colloids.~\cite{Zengin2013e} Even though complex structures can be fabricated using lithographic techniques, and have already been used to promote plasmon-exciton coupling,~\cite{Bellessa2009, Schlather2013} this methodology is limited in the minimum feature size. In addition, metallic structures produced by lithography exhibit greater plasmon damping due to their surface roughness and inherent grain boundaries. All these aspects lower the quality factor ($Q$) and the near-field enhancement, decreasing the interaction strength one could potentially achieve with top-down fabricated structures.
To circumvent these limitations, one can resort to colloidal NPs, which are routinely synthesized in well-defined sizes and feature less ohmic losses due to their higher crystallinity as compared to top-down structures. Consequently, individual nanocrystals such as gold shells,~\cite{Fofang2011} silver rods,~\cite{Zengin2013e} and silver triangles~\cite{Zengin2015, Balci2013, DeLacy2015} have already been reported to display plexcitonic signatures in the presence of J-aggregates. These unmodified colloids, however, lack the ability to assemble into complex plasmonic designs that are required for specific applications.~\cite{Gonzalez-Ballestero2015}
More importantly, individual colloids do not take advantage of the additional field confinement that results from bringing together two or more closely-spaced NPs. In addition, silver colloids are known to oxidize, making the use of colloidal gold the preferred choice for plasmon-exciton systems. Very recently, plasmon-exciton coupling using individual molecules in combination with a nanoparticle-on-mirror (NPoM) configuration~\cite{Chikkaraddy2016}, as well as coupling between individual colloidal QDs with lithographically produced silver bow-tie antennas~\cite{Santhosh2016} have been reported, highlighting plasmonic cavities as promoters of strong light-matter interactions.\\
%
\begin{figure*}[t!]
\includegraphics[width=\textwidth]{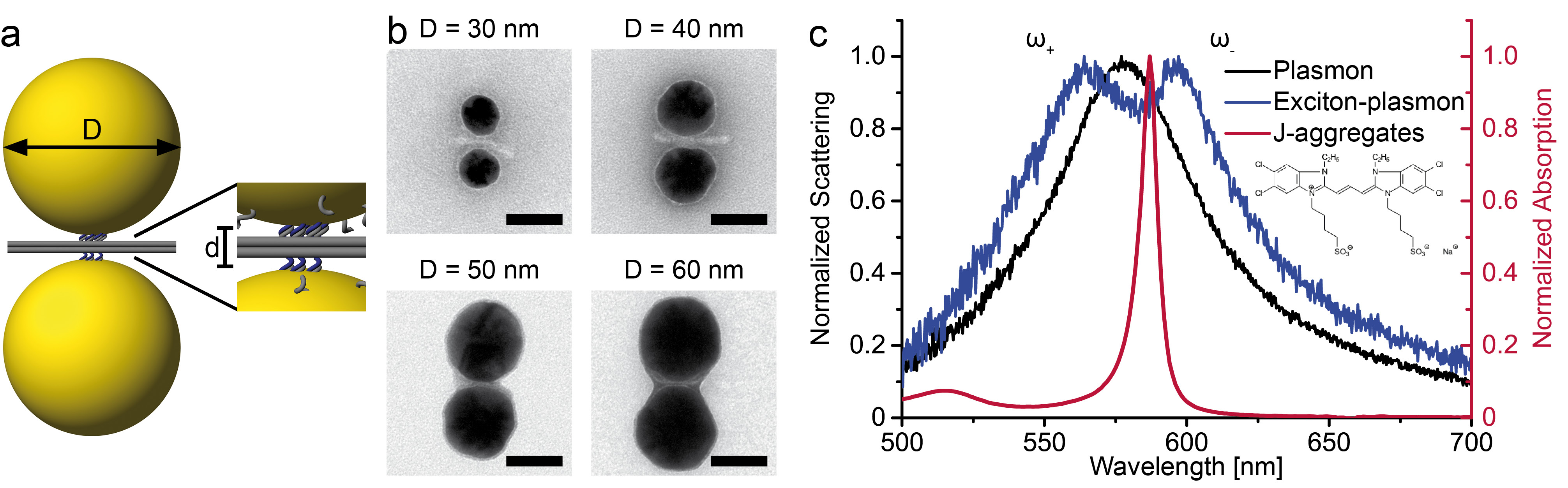}
\caption{
\textbf{(a)} Schematic of a 2-layer DNA-origami sheet templating a gold nanoparticle dimer with a designed separation of $\sim \, 5\,nm$. \textbf{(b)} TEM images of DNA sheets accommodating NPs of different diameters  ranging from $30\,nm$ to $60\,nm$. Scale bars are $40\,nm$. \textbf{(c)} Typical spectra of $40\,nm$ dimer with (dark-blue) and without (black) J-aggregates. J-aggregate absorption depicted in red. Inset shows chemical structure of the molecular exciton used in this work.}
\label{fig1}
\end{figure*}
%
DNA-origami is a technique routinely used to fabricate structures with nanoscale dimensions ($\sim\, 100\,nm$) and programmable designs.~\cite{Rothemund2006a, Douglas2009} In a one-pot reaction, a long viral single-stranded DNA (ssDNA) scaffold ($\sim\, 7$k\,bases) is folded by the help of $\sim\, 200$ complementary short 
synthetic ssDNA oligonucleotides. These structures can be used as templates with sequence-specific DNA binding sites, where nanocomponents functionalized with complementary DNA sequences can be attached to the binding sites (Figure~\ref{fig1}a).
~\cite{Schreiber2014, Zhang2015} 
DNA-templated metallic structures have already been tailored to affect the optical properties of nearby components such as custom-tuned ``hot spots'' for Surface Enhanced Raman Scattering (SERS),~\cite{Prinz2013, Kuhler2014, Thacker2014b, Pilo-Pais2014} enhancement and quenching of fluorophores~\cite{Acuna2012, Acuna2012a}, and colloidal quantum dots.~\cite{Ko2013, Samanta2014} In addition, the DNA-origami technique has been successfully used to tailor light, displaying strong circular dichroism~\cite{Kuzyk2012} as well as magnetic resonances.~\cite{Roller2015}\\
Here, we demonstrate strong-coupling between plasmons and excitons (J-aggregates) at room temperature and optical frequencies by exploiting the position accuracy that is achievable with the DNA-origami technique. This technique provides unprecedented control in the design of plexcitonic systems, bringing this technology one step closer to practical applications as compared to all previously proposed plexcitonic designs. By attaching pairs of colloidal gold nanocrystals to a DNA origami template, we fabricated a nanoantenna configuration with a fixed inter-gap distance of $\sim\,5\,nm$. The resonance frequency of the longitudinal plasmon mode of our constructs scales with the NP size and thus can be tuned across and even matched with the resonance of the desired exciton. This allowed us to observe normal mode splitting in the far-field scattering of individual constructs. \\\

Our DNA-templated nanodimer assemblies were fabricated using pairs of $30\,nm$, $40\,nm$, $50\,nm$, or $60\,nm$ diameter gold NPs functionalized with DNA linkers complementary to specific binding sites on a 2-layered DNA-origami sheet (Figure~\ref{fig1}a). Transmission electron microscopy (TEM) images of gel-purified structures reveal high yields (Supporting Figure~S1-S3) of correctly assembled particle dimers with designed interparticle gap of 5\,nm. For this particular gap size we find that dimers built from $40\,nm$ NPs are in closest spectral resonance with the exciton frequency of the cyanine-based dye used in this work (CAS\#~18462-64-1, FEW Chemicals GmbH). 
This methanol soluble dye readily stacks to form J-aggregates when dissolved in water. When absorbed to glass substrates, we find thin layers of J-aggregates to exhibit a scattering peak at $580\,nm$ ($2.14\,eV$) and a narrow FWHM linewidth of $30\,meV$ (Supporting Figure~S5b). For our measurements of combined plasmon-exciton systems, the assembled structures were deposited on a glass substrate and then immersed in a J-aggregate water bath solution ($50$\,$\mu M$). After overnight incubation, the samples were blown with nitrogen, flushing out most of the J-aggregate excess except at the location of the NP dimers. Far-field scattering measurements on individual structures were then performed using a home-built darkfield microscope (Supporting  Figure~S4 and Note~2). After recording the spectral response of the hybrid structures, the samples were exposed for 1 hour to continuous white light illumination under a 100\,x objective to completely photo-bleach the J-aggregates.~\cite{Zengin2015} This permitted us to additionally record the plasmon resonance of the structures without the contribution of the excitons.
Figure~\ref{fig1}c shows the far-field scattering spectra of a single AuNP dimer assembled using DNA-origami with J-aggregates before (dark-blue line) and after (black line) photo-bleaching the excitons. \\
Spectral red-shifts are more pronounced on dimers with bigger NP sizes due to stronger interparticle coupling. Thus, detuning of the plasmon mode with respect to the exciton resonance was achieved by building dimers with NPs sizes ranging from $30\,nm$ to $60\,nm$ (Figure~\ref{eq:2}a), while using the same origami design and thus a constant interparticle gap. This allowed us to tune the plasmon resonance wavelength between $2.05\,eV$ and $2.20\,eV$ across the exciton resonance at $2.14\,eV$. As a result of the coupling, the scattering spectrum splits into hybrid states of lower ($\omega_-$) and higher ($\omega_+$) energies. 
Polarization-resolved measurements show that only the longitudinal mode couples, as only this mode matches the exciton resonance (Supporting  Figure~S5b). 
%
\begin{figure*}[t!]
\includegraphics[width=\textwidth]{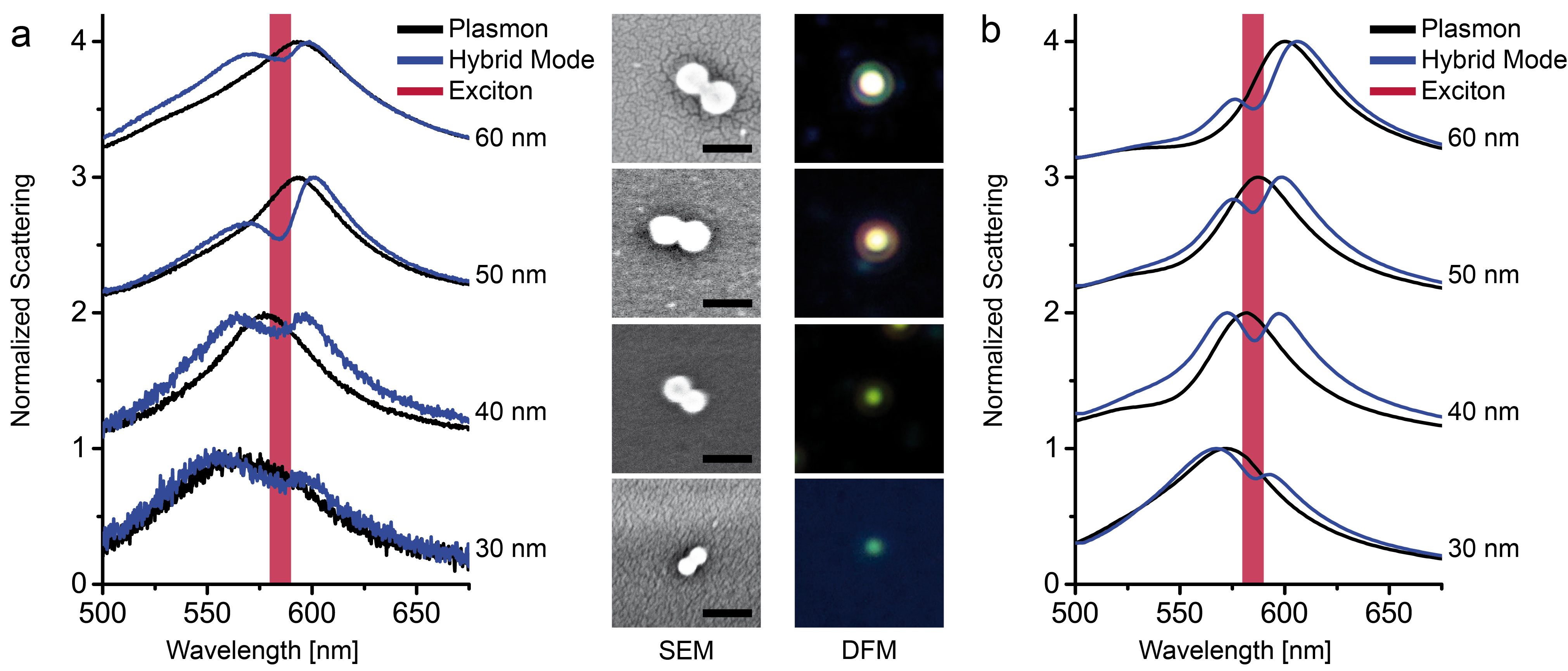}
\caption{
\textbf{(a)} Normalized scattering spectra before (dark-blue) and after (black) photo-bleaching the J-aggregate for $30\,nm$, $40\,nm$, $50\,nm$ and $60\,nm$ NPs dimers. The spectral region shaded in red covers the energy resonance of the J-aggregate. The right panels show the structures corresponding to each spectrum under SEM and darkfield microscopy. Scale bars: $100\,nm$. \textbf{(b)} Numerical simulations show excellent agreement with the experimental data.}
\label{fig2}
\end{figure*}
%
For each particle size, we measured the scattering spectra of several individual structures and present exemplary spectra for all sizes in Figure~\ref{fig2}a. Numerical simulations show excellent agreement with the experimental data (Figure~\ref{fig2}b, Supporting Note~4). Using the data collected for Figure~\ref{fig2}a, we followed the position evolution of the upper $\omega_+$ and the lower $\omega_-$ hybrid states as a function of the longitudinal plasmon mode and NP radius ($R$). As expected, the energy positions display a pronounced avoided crossing, characteristic of strong coupling (Figure~\ref{fig3}).~\cite{Torma2015}\\\
 
The Rabi frequency ($\Omega$) corresponds to the spectral separation of the normal modes ($\Delta\omega)$ when the plasmon and the exciton are at perfect resonance. To extract its value, we modeled the system as a two coupled harmonic oscillators with complex frequencies $\tilde{\omega} = \omega + \imath \Gamma/2 $. The resulting complex eigenvalues are~\cite{Gomez2014}
\begin{equation}
\tilde{\omega}_{\pm} = \dfrac{\tilde{\omega}_{p} + \tilde{\omega}_{qe}}{2} \pm \sqrt{g^2 + \dfrac{(\tilde{\omega}_{p} -\tilde{\omega}_{qe})^2}{4}}, \label{eq:1}
\end{equation}
\noindent where $\tilde{\omega}_{p}$ and $\tilde{\omega}_{qe}$ are the complex frequencies of the plasmon and quantum emitter \mbox{(J-aggregate)}, respectively, and g is the coupling constant. The position of the hybrid modes ($\omega_\pm$) as well as the FWHM line-widths of the exciton ($\Gamma_{qe}$) and the plasmon ($\Gamma_{p}$) were extracted by Lorentzian fitting of the corresponding scattering spectra, as described in Supporting Note~3. The spectral separation between the upper and lower modes resonances, $\Delta\omega$, is then given by (Supporting  Note~3):
\begin{equation}
(\Delta\omega)^2 = \sqrt{(\omega_p-\omega_{qe})^2(\Gamma_p-\Gamma_{qe})^2 + \left(4g^2+(\omega_p-\omega_{qe})^2-\frac{(\Gamma_p-\Gamma_{qe})^2}{4}\right)^2}.\label{eq:2}
\end{equation}
Equation \ref{eq:2} reduces to the commonly used Rabi Splitting $\Omega = 2 \sqrt{g^2 - (\Gamma_{p} -\Gamma_{qe})^2/16}$
%
\begin{figure}[t!]
\centering
\includegraphics[width=\textwidth]{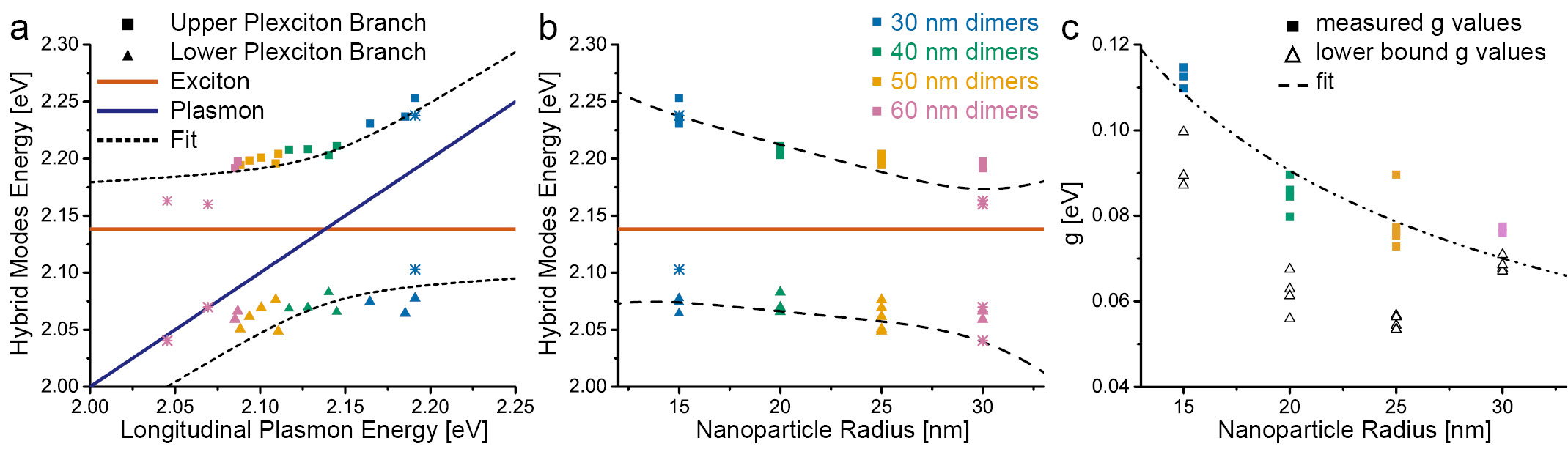}
\caption{
\textbf{(a)} Hybrid plasmon-exciton state energies are plotted as a function of their corresponding plasmon resonance and \textbf{(b)} as a function of the NP radius, showing the anticrossing behavior. Dashed lines are fits of the eigenvalues of a two-coupled harmonic oscillator with complex frequencies. Stars represent structures that do not fulfill the strong coupling condition and were not used in the fitting procedure. Both fittings (a) and (b) yield a Rabi splitting of $\Omega \sim\,150\,meV$. \textbf{(c)} Coupling constant (g)
values obtained for individual structures reveal a scaling of $g \propto 1/R^{\,n}$, $n = 0.63 \pm 0.08$ in very good agreement with the expected $g \propto 1/\sqrt{V_{eff}} = 1/R^{\,0.5}$. Dimers composed of smaller NP sizes display a larger coupling constant g. Triangles depict the lower bound for the strong coupling given by $g^2>(\Gamma_p^2+\Gamma_{qe}^2)/16$.}
\label{fig3}
\end{figure}
%
when $\omega_p = \omega_{qe}$, which sets a threshold of $g^2>(\Gamma_p-\Gamma_{qe})^2/16$ to ensure the splitting is real valued. An indication that the strong coupling regime has been reached~\cite{Torma2015} is given by $g^2>(\Gamma_p^2+\Gamma_{qe}^2)/16$, which shows that the splitting between the new modes is greater than their linewidth. This sets a lower bound value of $g>60\,meV$ for the structures in resonance with the exciton ($\Gamma_{p,r=20nm}=240\,meV$). To extract the coupling constant $g$ of our system, we first performed a fitting on the upper $\omega_+$ and lower $\omega_-$ modes as a function of the plasmon frequency $\omega_p$ (Figure~3a). This procedure assumes a constant $g$ for all particle sizes. 
The extracted exciton-plasmon coupling $g_{fit}$ is $\sim\,90\,meV$, which results in a Rabi splitting of $\Omega \sim\,150\,meV$. \\
To account for the varying $g(R)$ as the NP radius is changed, we then fitted the exponential function $g = a*R^{\,n}$ (Figure~3c) using radial-dependent parameters ($\Gamma_{p\,(R)}, \omega_{p\,(R)}$) extracted from the recorded data (Supporting Figure~S6). Here, $a$ and $n$ are fitting parameters. This analysis revealed a coupling constant which scales with the NP radius as $g \sim\, 1/R^{\,n}$, $n = 0.63 \pm 0.08$, showing that higher coupling constants are obtained for smaller NP sizes. At the expected anticrossing position ($R = 20\,nm$), both fitting procedures provide an equal value of $g \sim\,90\,meV$. Note that in our analyses only those spectra with a real valued coupling constant were taken into account when considering equation~\ref{eq:2}. Structures with values far from resonance (depicted with asterisks in Figure~\ref{fig3}) only exhibited a Fano-like signature or absorbance dip enhancement.
Figure 3c displays the $g$ dependence on the radius and clearly shows that dimers with reduced NP size exhibit larger coupling constants. The g values were extracted by replacing the individual $R, \omega_p, \Delta\omega, \Gamma_p$ parameters in equation~\ref{eq:2} for each measured NP-dimer.
Following reference~32, we approximated the effective mode volume of two closely spaced NPs to be a cylinder with a circular base of diameter $\sqrt{R\,d}$ (width of the induced surface charge) and height $d$ (gap distance), $V_{eff} \propto R\, d^{\,2}$.\,~\cite{Savage2012}
The measured $g \sim\, 1/ R^{\,n}$ with $n = 0.63 \pm 0.08$ is in very good agreement with the expected scaling of $g \propto 1/\sqrt{V_{eff}} \sim 1/R^{\,0.5}$.
In summary, we observe that smaller particles exhibit the strongest coupling. However, the steep decrease in scattering of smaller NP systems makes the study of individual structures below $30~nm$ in size challenging. Moreover, we also expect that as the NP size is further reduced, surface scattering damping would start to dominate. Thus, there is an ideal NP size where the coupling strength is maximal. This regime was not accessible with our current experimental setup.\\\

%

We have successfully demonstrated that DNA templates can be used to rationally engineer plexcitonic systems that display hybridized modes between plasmons and molecular excitons (J-aggregates) at room temperature. 
Our structures are programmed to self-assemble in solution and take full advantage of the field confinement produced by closely-spaced metallic colloidal nanocrystals. The coupled plasmon mode can be custom-tuned to be in resonance with the exciton of interest by setting the desired interparticle separation and nanoparticle size. 
Moreover, one could further exploit the full addressability of the DNA-origami technique to incorporate and precisely position additional nanocomponents, such as individual dyes or quantum dots. As such, the DNA-origami technique provides an unparalleled control in the fabrication of plexcitonic systems, and represents a promising platform to achieve fully integrated nanobreadboards and quantum nanocircuits. In future work, we will further investigate the coherent energy exchange between the plasmon and the exciton of our hybrid systems via second order photon correlation spectroscopy. The design flexibility and the parallel assembly formation of DNA-templates are ideally suited to study plasmon-exciton coupling and to fabricate complex structures for optical applications. 
\newpage

\singlespacing
\section{Associated Content}
\subsection*{Supporting Information Available} 
Supporting Information includes materials, detailed experimental methods, data analysis, and numerical calculation procedures. This material is available free of charge via the Internet at http://pubs.acs.org.
\section{Author Information}
\subsection{Corresponding Author}
*Email: m.pilopais@lmu.de
\subsection*{Author Contributions} 
E.M.R, A.H., T.L., and M.P conceived the experiment. E.M.R and M.P. conducted the experiments and analyzed the results. C.A. performed the numerical calculations. All authors interpreted the data and reviewed the manuscript.
\subsection*{Notes} 
The authors declare no competing financial interests.
\section*{Acknowledgments} 
This work was funded by the Volkswagen Foundation, the DFG through the Nanosystems Initiative Munich (NIM), and the ERC through the Starting Grant ORCA. A.H. acknowledges funding by the ERC starting grant No.\,336749. C.A. would like to acknowledge support by the Office of Research and Economic Development at University of Nebraska Lincoln and the NSF Nebraska MRSEC.
\singlespacing
\bibliography{bibliography}

\newpage
\section*{For TOC Only}
\vspace{50pt}
\begin{figure}[H]
\centering
\includegraphics[scale=1]{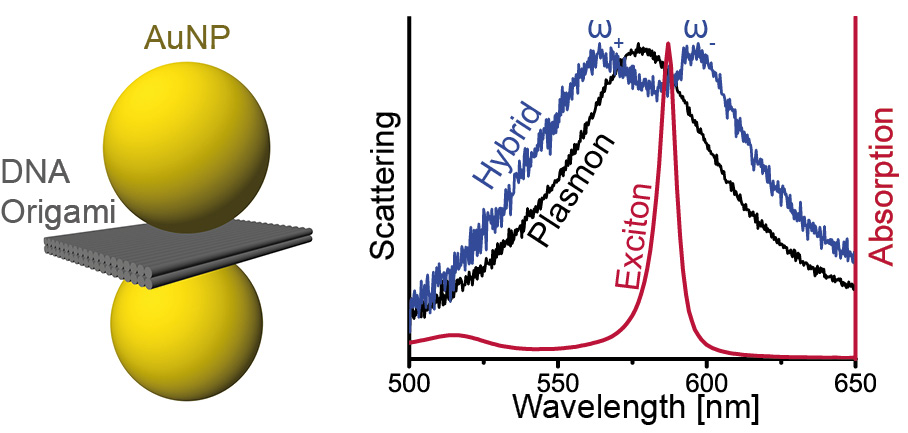}
\end{figure}
\end{document}